\documentclass[%
preprint,
 amsmath,amssymb,
 aps,
]{revtex4-2}

\usepackage{graphicx}
\usepackage{dcolumn}
\usepackage{bm}
\usepackage[mathlines]{lineno}

\begin{document}

\preprint{APS/123-QED}

\title{Ion Reflection by a Rippled Perpendicular Shock}

\author{Yuri V. Khotyaintsev}
\email{yuri@irfu.se}
\author{Daniel B. Graham}%
\affiliation{%
 Swedish Institute of Space Physics, Uppsala
}%


\author{Andreas Johlander}
\affiliation{%
 Swedish Institute of Space Physics, Uppsala
}%
\affiliation{
 Swedish Defence Research Agency, Stockholm
}%

\date{\today}

\begin{abstract}
We use multi-spacecraft Magnetospheric Multiscale (MMS) observations to investigate electric fields and ion reflection at a non-stationary collisionless perpendicular plasma shock.
We identify sub-proton scale (5-10 electron inertial lengths) large-amplitude normal electric fields, balanced by the Hall term ($\mathbf{J} \times \mathbf{B}/ne$),
as a transient feature of the shock ramp related to non-stationarity (rippling).
The associated electrostatic potential, comparable to the energy of the incident solar wind protons, decelerates incident ions and reflects a significant fraction of protons, resulting in more efficient shock-drift acceleration than a stationary planar shock.
\end{abstract}

\maketitle


\emph{Introduction}. Collisionless shocks are some of the most important particle accelerators in the Universe \citep{Treumann2009}. Particles can reach high energies, which requires interaction with a large shock over a long time, as in the supernova shocks \citep{reynoso2013, vink2022}. However, the initial energization of the cold upstream plasma happens in the narrow shock transition region, which can be studied using in-situ data at Earth's bow shock.

In collisionless plasmas, Coulomb collisions are rare, which necessitates alternative dissipation mechanisms to maintain the shocks. Super-critical shocks achieve this through plasma reflection,
which strongly depends on the shock geometry. 
For quasi-perpendicular geometry, when the angle between the upstream B and the shock normal $\theta_{Bn}>45^{\circ}$, the reflected ions reach the downstream after a few gyrations in the upstream, providing the major contribution to the increase of the downstream ion temperature \cite{sckopke1983}. 

The reflection depends on the detailed structure of electromagnetic fields (including waves) and the feedback of the reflected ions on the shock structure and is generally more complex than specular reflection \cite{sckopke1983}. 
Ramps of quasi-perpendicular shocks often contain complex large-amplitude electric fields\citep{bale2007,goodrich2019}. To evaluate the effect of such fields on particles, it is necessary to know their scale and potential, which are difficult to estimate.
Typically, one had to assume the structures propagate at the average speed of the ramp \citep{walker++04}, which is not valid for non-stationary shocks \citep{johlander2023}. For super-critical quasi-perpendicular shocks, the non-stationarity is related to ion reflection and results in shocks ripples and reformation\citep{biskamp1972,lembege1992,hellinger2002,burgess2016,lobzin2007, umeda2018}. The ripples correspond to ion-scale shock micro-structure, which has been resolved using the high-cadence in-situ measurements by Magnetospheric Multiscale (MMS) mission\citep{johlander2016,gingell2018,johlander2018, madanian2021}. Here, we investigate the ion reflection in a rippled perpendicular super-critical shock.





\begin{figure}
\includegraphics[width=8.6cm]{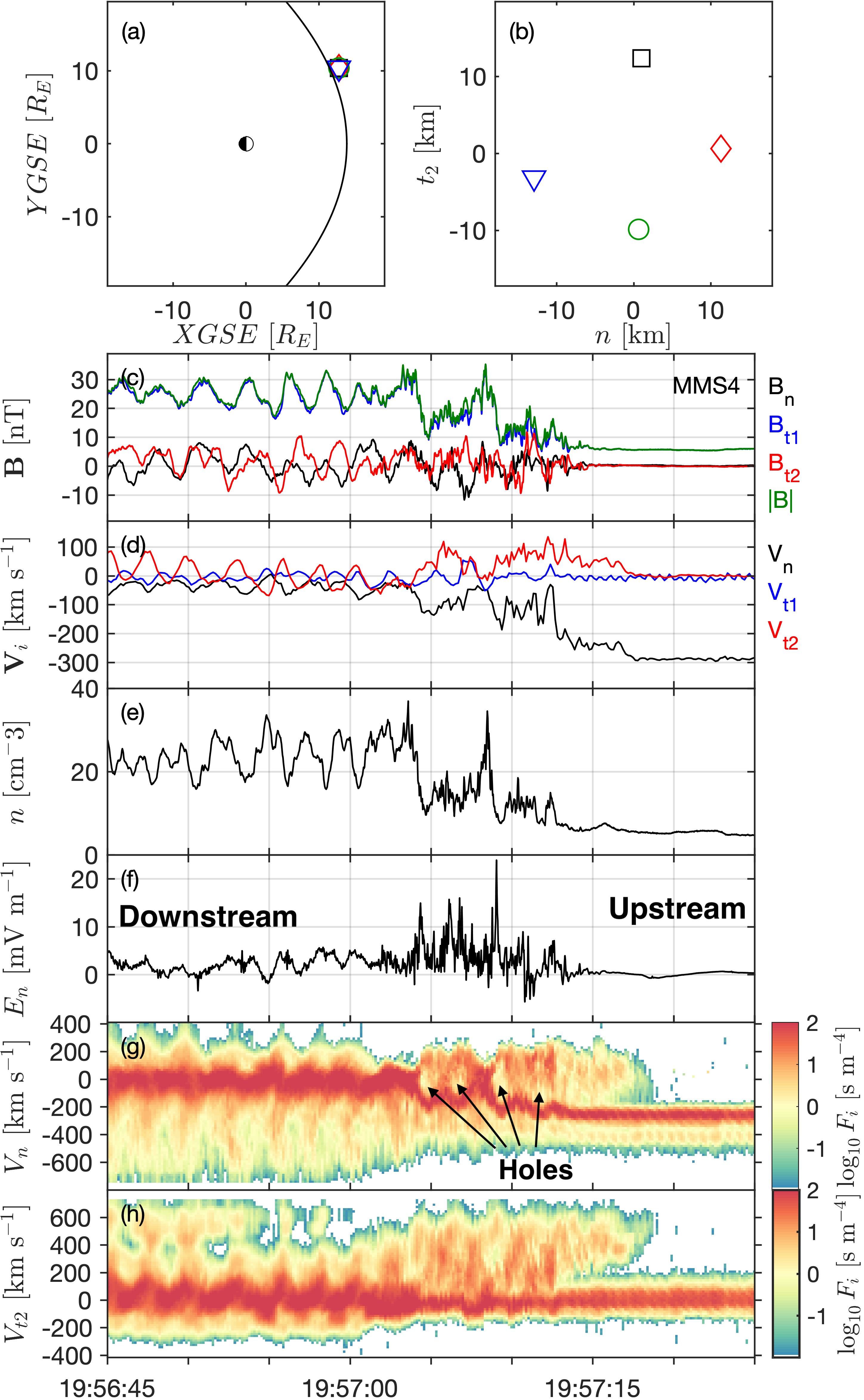}
\caption{Overview of shock on November 14, 2017. (a) Location of the MMS spacecraft with respect to the model bow-shock, (b) MMS tetrahedron configuration in the $\mathbf{\hat{n}}$-$\mathbf{\hat{t}_2}$ plane, (c) magnetic field from FGM \citep{russell2014}, (d) ion velocity in the NIF shock frame and (e) electron density from FPI, (f) $E_{n}$ measured by EDP\citep{lindqvist2014,ergun2014} low-pass filtered at 15 Hz (to remove high-frequency electrostatic fluctuations), (g) and (h) 1D reduced ion VDFs (integrated over the two other dimensions) in the $\mathbf{\hat{n}}$ and $\mathbf{\hat{t}_2}$ directions (in the NIF shock frame). Panels (c) to (h) show data from MMS4.
\label{fig:over}}
\end{figure}

\emph{Observations}. We analyze a perpendicular bow shock, $\theta_{Bn}\sim$ 89$^{\circ }$, observed by MMS on November 14, 2017, see Fig.~\ref{fig:over}. The shock crossing is from down- to upstream. The four spacecraft are separated by $\sim$10 km, shown in Fig.~\ref{fig:over}b, and the shock parameters are summarized in Table~\ref{tab:params}, which correspond to rather standard conditions for the Earth's bow shock \citep{lalti2022}.
The shock normal $\mathbf{\hat{n}}$=[0.92 0.37 0.10] in geocentric solar ecliptic system (GSE) is determined by the mixed mode method \citep{abraham-shrauner1972} using average parameters over the upstream (19:57:23--19:57:32~UT) and downstream (19:56:13-19:56:36~UT) intervals.
Fig.~\ref{fig:over}c shows the magnetic field from MMS4 in a $\mathbf{\hat{n}}$, $\mathbf{\hat{t}_1}$, $\mathbf{\hat{t}_2}$ system, where $\mathbf{\hat{t}_2} = \mathbf{\hat{n}} \times \mathbf{B_u}/|\mathbf{\hat{n}} \times \mathbf{B_u}|$ ($B_u$ is the upstream B) and $\mathbf{\hat{t}_1} = \mathbf{\hat{t}_2} \times \mathbf{\hat{n}}$. The shock normal velocity $V_{sh}=-57 \pm 20$~km/s (mean and standard deviation over the downstream interval) is determined from mass flux conservation. We use $V_{sh}$ to transform the data into the normal incidence (NIF) shock rest frame. We observe large variations in $B$, electron density $N_e$, and $V_{i,n}$, and large amplitude spikes in $E_n$ within the shock transition. Specifically, two sharp increases in $B$ and $N_e$ correspond to the shock ramp encounters, indicating the presence of the shock ripples. The NIF rest frame is the average shock frame, and ripples cause an additional smaller-scale motion of the ramp in this frame. 

\begin{table}[b]
\caption{\label{tab:table1}%
Shock and upstream plasma parameters.
}
\begin{ruledtabular}
\begin{tabular}{lc}
\textrm{Parameter}&
\textrm{Value}\\
\colrule
Magnetic field magnitude $|B_u|$ & 6 nT\\
SW density $n_{sw}$ & 4.7  cm$^{-3}$\\
SW velocity $V_{sw}$ & 390 km s$^{-1}$\\
Alfv\'en Mach number $M_A$ & 6.4\\
Magnetosonic Mach number $M_{ms}$ & 5.3\\
SW ion $\beta_{i,u}$ & 0.2\\
Shock normal in GSE $\mathbf{\hat{n}}$ & [0.92 0.37 0.10]\\
$\theta_{Bn}$ & 89$^{\circ }$
\label{tab:params}
\end{tabular}
\end{ruledtabular}
\end{table}

To characterize the transition region, we plot the reduced 1D ion velocity distribution functions (VDFs) as a function of $V_n$ and $V_{t2}$ (Fig.~\ref{fig:over}g, h). The incoming solar wind proton beam is the dominant component (after 19:57:15~UT). The other lower density and faster (observed at $V_n \sim$ -400 km/s) component is He$^{++}$ (discussed later). Between 19:57:03 and 19:57:08~UT, the proton beam slows down in several steps, and reflected protons ($V_n >0$, moving upstream) appear. The main beam has a constant $V_{t2} \sim$0. The reflected protons have $V_{t2}>0$. As a result of acceleration by the upstream convection electric field $\mathbf{E_u}$ (aligned with $\mathbf{\hat{t}_2}$), the reflected protons have higher energy than the ions directly transmitted downstream, and their gyration in the downstream results in the characteristic vertical stripes (panels g, h before 19:57:02~UT). Such gyrating protons constitute $\sim$8\% of the downstream ion density, which we obtain by integrating the VDF (averaged between 19:56:45 and 19:57:00~UT) and excluding the transmitted population defined by plasma-frame velocities below $2 \times V_{Tp\mathrm{min}}$, where $V_{Tp,\mathrm{min}}$ is the thermal velocity corresponding the lowest value of the temperature $T_{p,\mathrm{min}}\sim$50~eV observed during this interval.  

We identify several holes in the ion phase space (Fig.~\ref{fig:over}g, between 19:57:03 and 19:57:08~UT) having similar time scales to $B$ and $N_e$ variations. The holes originate from the oscillation of the ion reflection point (shock ramp) relative to the spacecraft; consequently, the spacecraft moves between the foot, where the incoming and reflected beams are well separated (middle of a hole), and the downstream, where the two beams merge (edge of a hole). The presence of such holes is a characteristic signature of the shock ripples \citep{johlander2016,johlander2018}.

\begin{figure}
\includegraphics[width=8.6cm]{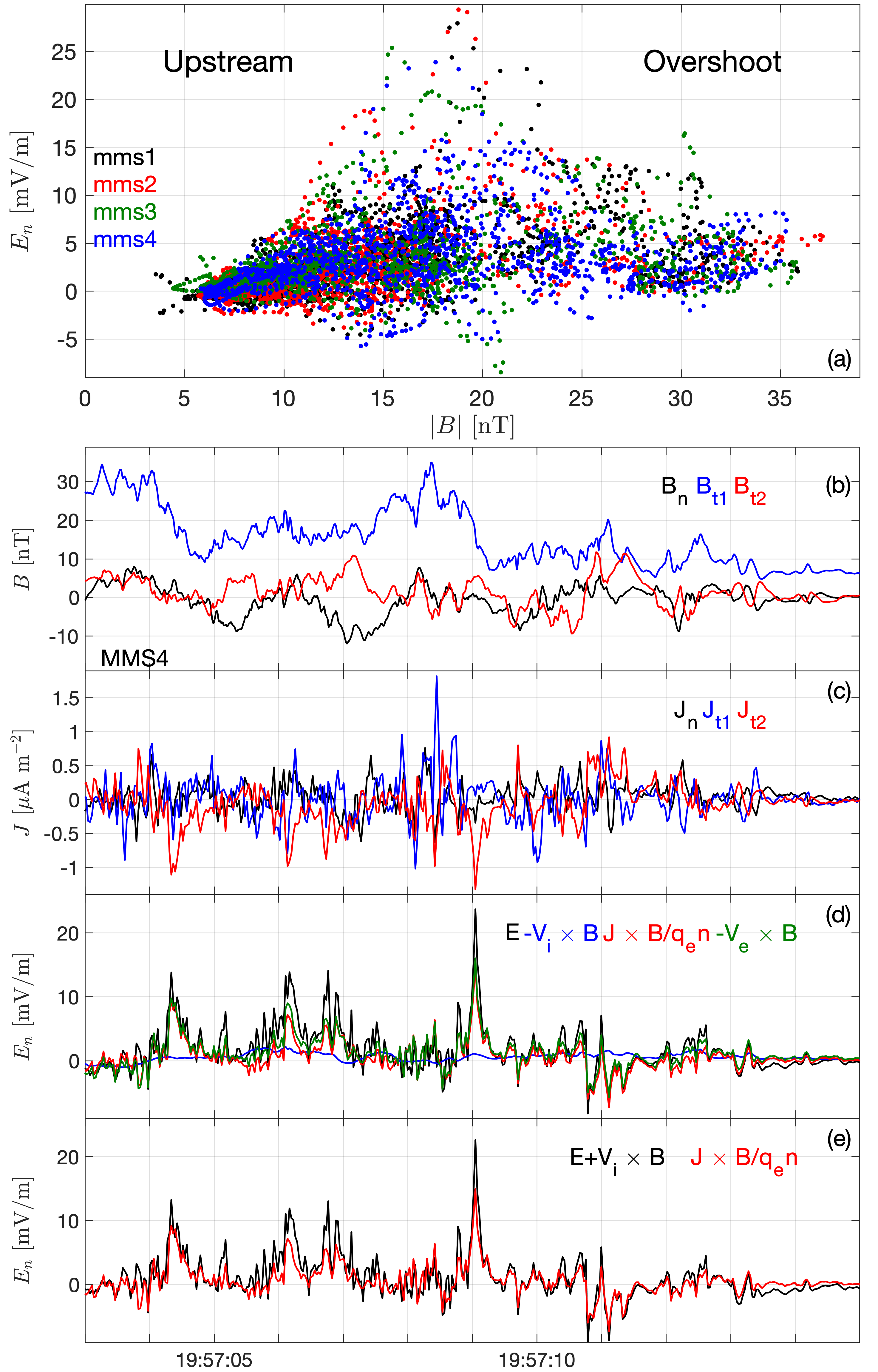}
\caption{Electric field and force balance normal to the shock. (a) $E_n$ from the four MMS spacecraft as a function of $|B|$. Panels (b) to (e) show $\mathbf{B}$, $\mathbf{J}$ (computed from the particle moments), $E_n$, and the terms in Eq.~\ref{eq:gol} from MMS4 in NIF shock frame.
\label{fig:ohmslaw}}
\end{figure}

We investigate whether the large amplitude spikes in $E_n$ (Fig.~\ref{fig:over}f) represent a permanent feature of the ramp using $|B|$ as a proxy of the spacecraft location within the ramp (Fig.~\ref{fig:ohmslaw}a): 
small $|B| \sim $6~nT corresponds to the upstream, and large $|B| > $30~nT corresponds to the overshoot/downstream. 
For a permanent feature, the spikes would be present for both the inward and outward crossings of the ramp. However, this is not the case and, generally, $E_n \sim 0$ (slightly positive) for all $|B|$, which suggests the spikes are a transient feature.
One example is the spike at 19:57:09 UT (Fig.~\ref{fig:ohmslaw}d) coinciding with an interval of decreasing $|B|$ (outbound crossing) at $|B|\sim$20~nT. If the spike were a permanent feature, it would also be present at $|B|\sim$20~nT during increasing $|B|$ (inbound crossing) at 19:57:07.2--08.2 UT, but this is not observed. 
We have inspected all the major $E_n$ spikes observed by the four spacecraft and found that most are observed for decreasing $|B|$ (outbound crossing of the ramp). This is consistent with super-imposed epoch analysis for a different rippled shock with similar $M_A$ analyzed in Ref.~\citep{johlander2018}, see Supplementary material. We conclude that the $E_n$ spikes are transient and are located at the particular phase of the ripple when the ramp speed (in the average shock frame) is negative (towards the downstream).

\begin{figure}
\includegraphics[width=8.6cm]{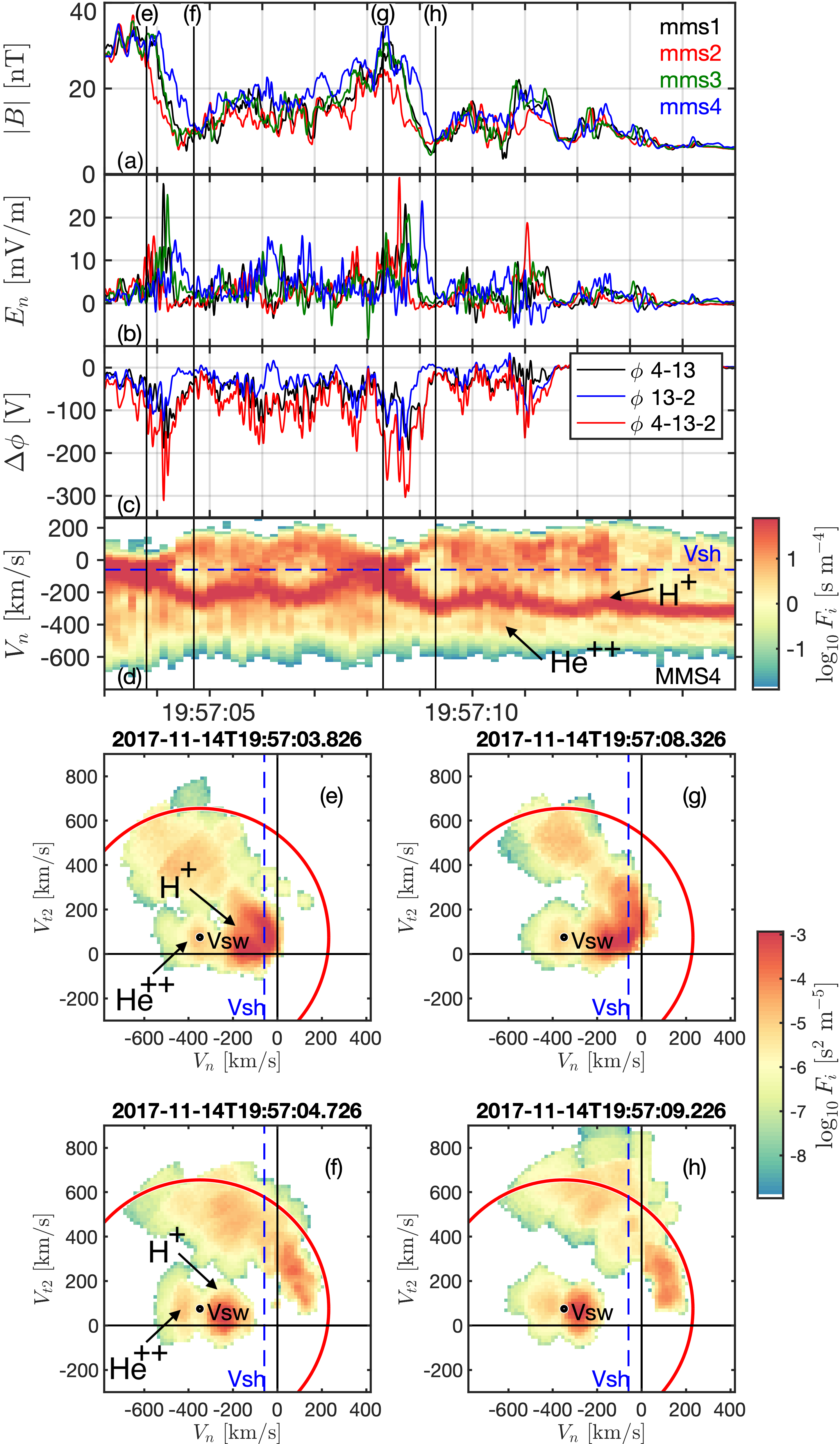}
\caption{Cross-shock potential and ion reflection. Panels (a) to (d) show $\mathbf{B}$, $E_n$ from the four spacecraft, cross-shock potential $\Delta \phi$ from integration over 2 points MMS 4-13, 13-2, 3 points MMS 4-13-2, and reduced 1D ion VDF from MMS4. Panels (e) to (h) show reduced 2D ion VDF at four different times marked by vertical lines in panels (a-d).
\label{fig:reflection}}
\end{figure}

We investigate the normal force balance at the shock using the generalized Ohm's law \citep{khotyaintsev2006}, neglecting electron inertia: 
\begin{equation}
\mathbf{E} + \mathbf{V_i} \times \mathbf{B} = \frac{1}{ne} \mathbf{J} \times \mathbf{B} 
- \frac{1}{ne} \nabla p_e.
\label{eq:gol}
\end{equation}
 Neglecting the minor contribution from the pressure term (estimated using multi-spacecraft data, not shown), we see in Fig.~\ref{fig:ohmslaw}e that Ohm's law is satisfied, confirming that the different terms are measured accurately. The electrons are well magnetized, $E_n \sim -(\mathbf{V_e} \times \mathbf{B})_n$, but ions are not $\mathbf{V_i} \times \mathbf{B} \sim 0$. The spikes in $E_n$  are primarily balanced by the Hall term, $E_n \sim (\mathbf{J} \times \mathbf{B}/ne)_n$, which indicates that they have a sub-proton scale. 
This is consistent with the multi-spacecraft data in Fig.~\ref{fig:reflection}a,b.  
The large differences in $|B|$ observed at different spacecraft (times e-f, g-h) indicate that 
the ramp thickness, $L_r$, is comparable to the spacecraft separation $\sim$23 km $\sim 9 d_e$ $\sim$0.23~$d_{iu}$. Ramp crossings (the largest changes in $|B|$) and $E_n$ spikes have similar scales, confirming the spikes' sub-proton $\sim L_r$ scales. Spikes of similar amplitude are observed at four spacecraft at different times (Fig.~\ref{fig:reflection}b), indicating the spikes exist for at least $\sim$0.6~s.  

Comparing $E_n$ to the ion VDFs in Fig.~\ref{fig:reflection}d, we find that the $E_n$ spikes coincide with locations where the incoming ion beam slows down. 
The positive $E_n$ spikes correspond to $\mathbf{E}$ pointing upstream, decelerating the upstream ions. To evaluate the effect of $E_n$ on the ions, we estimate the change in the electrostatic potential, $\Delta \phi = -\int E_n dn$. 
Integration in time using the average shock speed, $dn=V_{sh} dt$, would yield erroneous results as the local shock speed is highly variable due to the ripples (\cite{johlander2023}). Instead, 
 we utilize the spacecraft separation in the $\mathbf{\hat{n}}$-$\mathbf{\hat{t}_2}$ plane (Fig.~\ref{fig:over}b) for integration in space: 
we average $E_n$ from MMS1 and MMS3, which are located at the same position in $\mathbf{\hat{n}}$ so that we get three measurements, $E_{n2},E_{n13},E_{n4}$, separated along $\mathbf{\hat{n}}$ by a distance similar or below the scale of the $E_n$ spikes. 
We use these three points for integration. 
The resulting potential is shown in Fig.~\ref{fig:reflection}c. The potential differences are negative, corresponding to $\Delta \phi$ increasing towards the downstream, so that the positive ions can reflect off the potential barrier if the barrier is higher than the ion kinetic energy. The potential drop between the individual spacecraft pairs reaches 150~V and 300~V across the constellation; the overall $\Delta \phi$ is likely somewhat higher than 300~V as the MMS constellation does not cover the entire ramp. 
The observed $e\Delta \phi$ values are comparable to the average upstream H$^+$ kinetic energy $\sim$400~eV and thus can decelerate and reflect a significant fraction of H$^+$.

We examine the 1D ion VDF in Fig.~\ref{fig:reflection}d, focusing on the two intervals containing the largest $\Delta \phi$ (marked vertical lines e-f and g-h). Both intervals are related to ion phase-space holes. On the upstream (right-hand) side, the intervals end at the centers of the holes (lines f, h) where we see incident ($V_{n}\sim -300$~km/s) and reflected ($V_{n}\sim +200$~km/s) beams. On the downstream (left-hand) side, the intervals start (lines e, g) at the reflection point, $V_{n}\sim -50$~km/s $\sim V_{sh}$. Spikes in $E_n$ are located close to the reflection point on one side of the hole, i.e., when the ramp moves outward. Across each interval, the incoming proton beam is decelerated by $\Delta V_{\mathrm{p} n}\sim 250$~km/s. This corresponds to a change in proton kinetic energy in the normal incidence shock rest frame of $\sim$330~eV, comparable to deceleration due to the observed potential change, $e\Delta \phi \gtrsim$300~eV. Thus, the observed deceleration and reflection can be fully explained by the strong $E_n$.

We show samples of 2D VDFs in the $\mathbf{\hat{n}}$-$\mathbf{\hat{t}_2}$ plane at times (e)-(h) in the bottom panels. At the reflection point (times e, g),  most of the protons are at $V_n \sim V_{\mathrm{sh}}$, $V_{t2}\sim 0$. We do not observe ions with $V_{t2}<0$; $V_{t2}<0$ is expected from cyclotron turning in the downstream $\mathbf{B}$.
In the center of the hole (times f, h) we identify the incoming H$^{+}$ beam slightly decelerated relative to the upstream $V_{\mathrm{sw}}$ (black dot), as well as the gyrating  ($V_n <0$, $V_{t2}>0$) and reflected ($V_n >0$, $V_{t2}\sim 0$) components.  The circles show velocities corresponding to $| \mathbf{V} - \mathbf{V_{sw}} | = 2 | \mathbf{V_u} \cdot \hat{n} |$, which assumes constant energy in the solar wind frame and specular reflection. The reflected component is close to this circle, indicating the ions experience nearly specular reflection. 
For the VDF at time (h), the reflected and returning components constitute  $\sim$27\% and $\sim$12\% of the incident proton density, respectively. For a uniform time-stationary shock, these will have equal densities. We interpret this difference in densities as a more efficient local reflection compared to a location from where the gyrating protons originate. We also note that the fraction of reflected protons is higher than the average downstream fraction of gyrating protons.

In contrast to H$^{+}$, the alphas (He$^{++}$) are decelerated by $E_n$ but not reflected. They have a double mass-to-charge ratio and would require the double potential barrier to be reflected. The incoming alpha beam is seen in Fig.~\ref{fig:reflection}d at $V_n\sim$-500~km/s. 
The alpha beam is seen in both the hole center and the reflection point in the 2D VDFs, and it has a lower $V_{n} \sim V_{\mathrm{sw}n}$ at the reflection point (panels e, g). We note that FPI does not separate the different mass species, and for converting energy to ion velocities, we assume that all ions are protons; therefore, the plotted He$^{++}$ velocities are overestimated. Considering the correct mass-to-charge ratio for He$^{++}$, we find that alphas are decelerated by the energy corresponding to the potential drop $e\Delta \phi \sim$300~eV.


\emph{Discussion.} 
To impact the ion reflection, the $E_n$ spikes must exist at least at the time scale of reflection. For a constant $E_n$, this time can be estimated as $t_r\sim 4\times L_{r}/V_{uNIF}\sim$4$\times$23~km/290 km/s $\sim$0.3 s, which is smaller than the minimum lifetime of the observed spikes $>$0.6~s obtained from the multi-spacecraft observations and an order of magnitude smaller than the ripple temporal scale ($\sim$seconds $\sim$proton cyclotron period). So, at scales $\gtrsim t_r$, it is reasonable to treat ion reflection in a static shock structure. The observations strongly support this “static” picture of ion reflection; both the deceleration of the incoming H+ and He++ and the reflection of H+ are consistent with the interaction with the potential structure estimated from observed $E_n$. Also, the observations are in good agreement with a static 1D numerical model of the reflection (see appendix). 

The equation of motion for the normal component of the ion velocity is:
\begin{equation}
\frac{\mathrm{d}}{\mathrm{dt}} v_n = \frac{e}{m_i} E_n - \frac{e}{m_i} v_{t2}B,
\label{eq:ion_vn}
\end{equation}
and shows that ions can be reflected both by $E_n$ and the $\mathbf{v} \times \mathbf{B}$ terms, which are associated with two different scales: the scale of $E_n\sim L_r\sim$23~km and the convective gyroradius the downstream field,  $\rho_{pc}\sim$115~km (for $B_d\sim$25~nT). In the absence of $E_n$, the incoming protons penetrate downstream for a distance $\sim\rho_{pc}$ experiencing cyclotron turning; however, if $E_n$ is sufficiently strong, the ions will decelerate over a shorter distance, diminishing the effect of $\mathbf{v} \times \mathbf{B}$ term. 
 Our observations supported by the numerical model show that the $E_n$, and not the $\mathbf{v} \times \mathbf{B}$ term, has the dominant contribution to the ion reflection in our event.

\emph{Conclusions.} We find that the strongest proton reflection comes from the sub-proton scale, $\sim$9 d$_e$, $E_n$ localized at the ramp of the rippled shock. This field is primarily balanced by the Hall term, $E_n \sim(\mathbf{J} \times \mathbf{B}/ne)_n$. The corresponding NIF potential of $\gtrsim$300 V is comparable to the energy of the incident solar wind protons, a significant fraction of which experiences specular reflection by $E_n$. At the same time, the heavier ions (He++) are decelerated and continue downstream. 
The large $E_n$ structures exist only at the particular phase of the ripple when the shock speed (in the average shock speed frame) is negative (towards the downstream). So, the resulting ion reflection is non-uniform and modulated by the ripples; we have a reflection of a large fraction of protons in some locations and a smaller fraction in others\citep{johlander2018}. This provides a specular reflection of a larger ion fraction than expected from a simple magnetic barrier, resulting in a more efficient shock drift acceleration.

\begin{acknowledgments}
This research was made possible with
the data and efforts of the Magnetospheric Multiscale mission team. 
This work is supported by the Swedish Research Council Grant 2018-05514
and the European Union's Horizon 2020 research and innovation program under grant agreement number 101004131 (SHARP) and the Swedish National Space Agency. The authors thank Dr. Ahmad Lalti for useful discussions.
\end{acknowledgments}

\appendix*

\begin{figure}
\includegraphics[width=8.6cm]{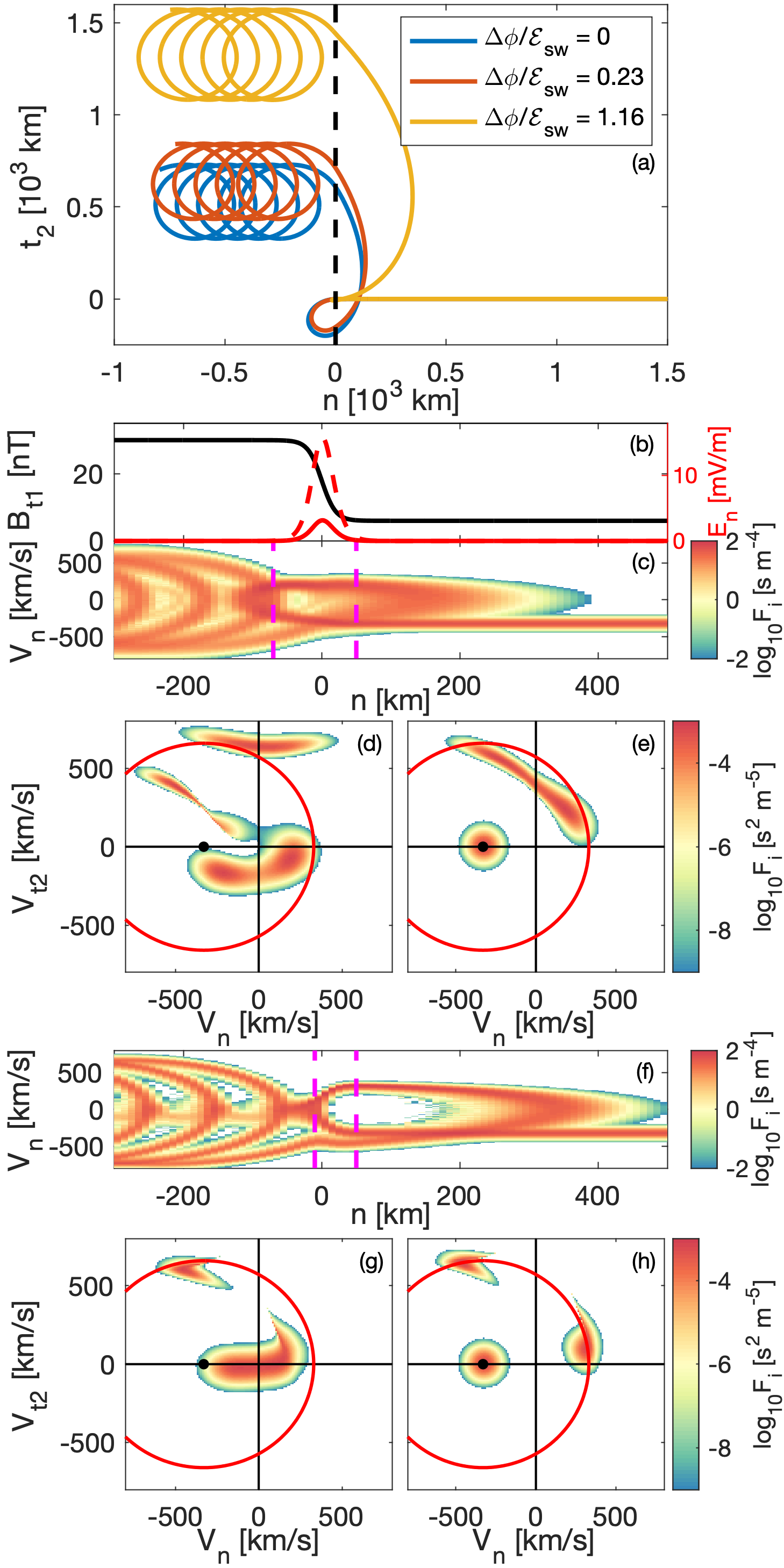}
\caption{Numerical model of ion distribution across a perpendicular shock. (a) Ion trajectories for different potentials. (b) Profiles of $B_{t1}$ and $E_n$. 
(c) Reduced ion distributions along $v_n$ for small $E_n$ (solid red line in panel b). (d) and (e) Reduced ion distributions in the $v_n$--$v_{t2}$ plane at the positions given by the magenta dashed lines in panel c. Panels (f)--(g) are ion distributions for strong $E_n$ (dashed red line in panel b) in the same format as panels c--e.
\label{fig:model}}
\end{figure}

\emph{Appendix on a model of ion reflection.} To illustrate the ion reflection in the presence of $E_n$, we employ a 1D numerical model describing ion motion in static E and B fields of the shock (see Supplementary material and Ref.~\citep{graham2024}). 
Fig.~\ref{fig:model}a shows particle trajectories of reflected protons for different values of $\Delta \phi$ relative to the upstream kinetic energy $\mathcal{E}_{sw}$. For $\Delta \phi/\mathcal{E}_{sw} < 1$ ions are decelerated by $\Delta \phi$ and cyclotron turned by the increase in $B_{t1}$ when reflected upstream, while for $\Delta \phi/\mathcal{E}_{sw} > 1$ ions are  specularly reflected by $\Delta \phi$. The reflected ions are accelerated by $E_{t2}$ in the upstream and transmitted across the shock. The resulting VDFs
are shown in Figs.~\ref{fig:model}c--\ref{fig:model}e and \ref{fig:model}f--\ref{fig:model}h for $\Delta \phi/\mathcal{E}_{sw} = 0.23$ and $\Delta \phi/\mathcal{E}_{sw} = 1.16$. For low $\Delta \phi/\mathcal{E}_{sw}$, ion reflection is due to the increase in $B_{t1}$ and results in a broad distribution of reflected protons (Fig.~\ref{fig:model}d). For large $\Delta \phi/\mathcal{E}_{sw}$, the reflected distribution is narrow in velocity space, and distinct populations are seen for the upward propagating reflected protons and the protons returning to the shock after reflection (Fig.~\ref{fig:model}h). Near the reflection point, the incoming distribution broadens along $V_n$, and a distribution of reflected protons returning to the shock is observed for large $V_{t2}$. These VDFs are similar to the observed VDFs in Fig.~\ref{fig:reflection}. Overall, the model distributions with large $\Delta \phi/\mathcal{E}_{sw}$ show better agreement with observations, suggesting that $\Delta \phi$ plays a crucial role in proton reflection. 

We note that this static 1D model does not explain the observed difference in the densities of reflected and returning protons, as modeling such behavior would require a time-dependent and/or 2D model.

%

\end{document}


\preprint{APS/123-QED}

\title{Supplemental material for ``Ion Reflection by a Rippled Perpendicular Shock''}

\author{Yuri V. Khotyaintsev}
\email{yuri@irfu.se}
\author{Daniel B. Graham}%
\affiliation{%
 Swedish Institute of Space Physics, Uppsala
}%


\author{Andreas Johlander}
\affiliation{%
 Swedish Instituite of Space Physics, Uppsala
}%
\affiliation{
 Swedish Defense Research Institute
}%

\date{\today}

\maketitle


\emph{Statistical map of E-normal}. To support the conclusion that the structure of the normal component of the electric is affected by presence of the ripples, we analyze a different quasi-perpendicular shock ($\theta_{Bn}$=65$^\circ$) with a similar Mach number ($M_A=7$) observed by MMS on January 6, 2016, at 00:33 UT reported in Ref.~\cite{johlander2018}. Similar to the shock on November 14, 2017 analyzed in the main text, this shock shows multiple clear holes in the ion phase space. Due to a very small velocity in the spacecraft frame, and the spacecraft spends a long time in the shock foot observing 10 periods of the ripple. Super-posed epoch analysis of the oscillation allowed to construct maps of $|B|$ and reflected ion density in the $\mathbf{\hat{n}} - \mathbf{\hat{t}_1}$ plane, see Fig.5a-b in Ref.~\cite{johlander2018}.

\begin{figure}
\includegraphics[width=8.6cm]{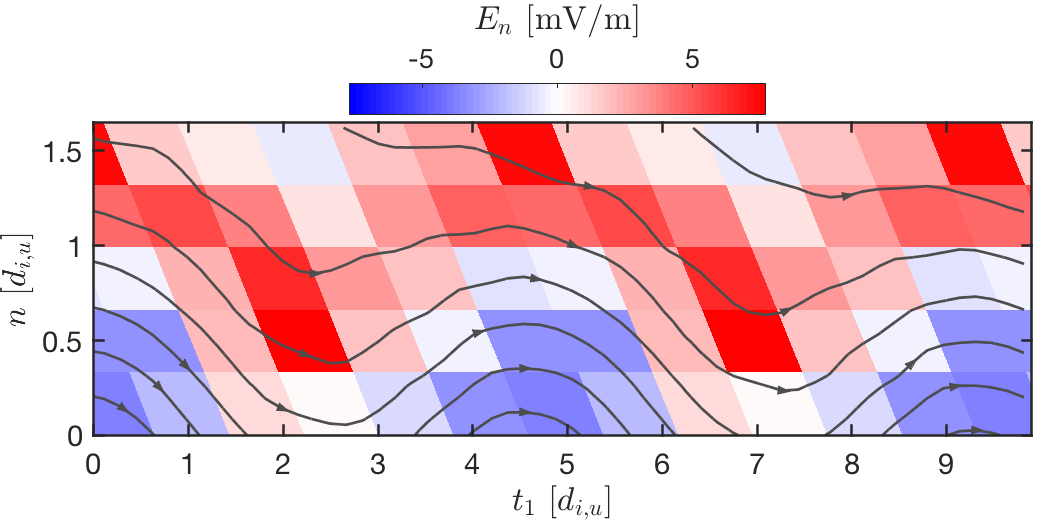}
\caption{2D histogram of $E_n$ for the rippled shock compiled with data from all four spacecraft E-field.  Approximate field lines derived from the magnetic field data are also shown.
\label{fig:e_n_super}}
\end{figure}

Here, we apply the super-posed epoch analysis to the normal component of the electric field, $E_n$. The result (shown in Fig.~\ref{fig:e_n_super}) reveals a distinct structure of $E_n$. Narrow spots of strong $E_n$ are localized to a particular phase of the ripple. As the ripple propagates in the $- \mathbf{\hat{t}_1}$ direction \cite{johlander2018}, the strong $E_n$ is located where the ramp (strong $|B|$ region) is moving in positive $\mathbf{\hat{n}}$ direction. This is consistent with the location observed for the perpendicular shock analyzed in the main text.

\emph{Liouville Mapping.}
To model the proton distributions across the shock, we use Liouville's theorem, which states that the phase space density is conserved along spacecraft trajectories. In the time-stationary case, this is expressed as 
\begin{equation}
f({\bf x}_1,{\bf v}_1) = f({\bf x}_0,{\bf v}_0),
\label{liovilleeq}
\end{equation}
where ${\bf x}_0$ and ${\bf v}_0$ are upstream solar wind position and velocity, and ${\bf x}_1$ and ${\bf v}_1$ are positions and velocities across the shock determined by equations of motion. 

To model the particle motion, we set up a simple bowshock model of a perpendicular shock in the normal incidence frame. The magnetic field and density profiles are given by 
\begin{equation}
B_{t1}(n) = - B_0 \mathrm{tanh}\left( \frac{n}{l} \right) + B_1, 
\label{Bmodeleq}
\end{equation}
\begin{equation}
n(n) = - n_0 \mathrm{tanh}\left( \frac{n}{l} \right) + n_1, 
\label{nmodeleq}
\end{equation}
where $l = 20$~km is the shock width, $B_0 = 12$~nT, $B_1 = 18$~nT, $n_0 = 13$~cm$^{-3}$, and $n_0 = 17$~cm$^{-3}$. 

The current density calculated from Ampare's law and Eq.~(\ref{Bmodeleq}) and given by
\begin{equation}
J_{t2}(n) = - \frac{B_0 \mathrm{sech}\left( \frac{n}{l} \right)^2}{\mu_0 l}. 
\label{Jmodeleq}
\end{equation}
The normal electric field is assumed to be Hall electric field term given by
\begin{equation}
E_n(n) = - \frac{J_{t2}(n) B_{t1}(n)}{e n(n)}. 
\label{Eneq}
\end{equation}
Equation (\ref{Eneq}) determines the cross-shock potential in the normal incidence frame. For the conditions given above, we calculate a shock potential of $130$~V. 
To consider the effect of different shock potentials, we multiply Eq.~(\ref{Eneq}) by a constant to change the potential. 
The electric field along the $t_2$ direction is given by
\begin{equation}
E_{t2}(n) = - \frac{V_{sw,n} (n_1 - n_0) B_{t1}(n)}{n(n)}, 
\label{Et2eq}
\end{equation}
where $E_{t2}$ yields ${\bf E} \times {\bf B}$ velocity equal to the bulk velocities upstream and downstream of the shock. 
Equations (\ref{Bmodeleq}), (\ref{Eneq}), and (\ref{Et2eq}) determine the ion trajectories across the shock. 

To perform the Liouville mapping, we assume an upstream ion distribution given by
\begin{equation}
f(v_n,v_{t2}) = \frac{n_i}{\pi v_i^2} \exp{\left( - \frac{(v_n - V_{sw,n})^2 + v_{t2}^2}{v_i^2} \right)}, 
\label{swdist}
\end{equation}
where $v_i = \sqrt{2 k_B T_i/m_i}$ is the proton thermal speed and $k_B$ is Boltzmann's constant. We assume the solar wind distribution is independent of position and time. For this model, there are no forces along the $t_1$ direction, so the two-dimensional reduced distribution in the $(v_n,v_{t2})$-plane is used. To model this shock we use $n_i = 4$~cm$^{-3}$, $T_i = 10$~eV, and $V_{sw,n} = -330$~km~s$^{-1}$. Reduced proton distributions in $(v_n,v_{t2})$-plane are computed across the shock by integrating backwards in time from each initial $(n,v_n,v_{t2})$. For trajectories where the particles reach upstream (1000~km upstream from the shock ramp) $f(n,v_n,v_{t2})$ is calculated using Eqs.~(\ref{liovilleeq}) and (\ref{swdist}). We use $f(n,v_n,v_{t2}) = 0$ for trajectories that never reach upstream of the shock ramp. Performing these calculations on a grid of $(n,v_n,v_{t2})$ enables the two-dimensional reduced distributions to be computed as a function of $n$ position. One-dimensional reduced distributions along $v_n$ and $v_{t2}$ are calculated by integrating the 2D distributions over $v_{t2}$ and $v_n$, respectively. 


%